\documentclass[twocolumn,showpacs,amsmath]{revtex4}
\usepackage{graphicx,amsmath}

\newcommand{\Journal}[4]{#1 \textbf{#2}, #3 (#4)}

\newcommand{\AgSbPbTe}{AgPb$_{m}$SbTe$_{2+m}$}

\begin{document}

\title{Resonant States in the Electronic Structure of the High Performance Thermoelectrics \AgSbPbTe\ ; The Role of Ag-Sb Microstructures }

\author{Daniel Bilc}
\author{S.D. Mahanti}
\affiliation{Department of Physics and Astronomy, Michigan State
University, East Lansing, MI 48824}
\author{Kuei-Fang Hsu}
\author{Eric Quarez}
\author{Robert Pcionek}
\author{M. G. Kanatzidis}
\affiliation{Department of Chemistry, Michigan State University,
East Lansing, MI 48824}

\begin{abstract}

  {\it Ab initio } electronic structure calculations  based on gradient corrected density functional theory were performed on a class of novel quaternary compounds \AgSbPbTe, which were found to be excellent high temperature thermoelctrics with large figure of merit {\it ZT}$\sim$2.2 at 800K.  We find that resonant states appear near the top of the valence and bottom of the conduction bands of bulk PbTe when Ag and Sb replace Pb. These states can be understood in terms of modified Te-Ag(Sb) bonds. Electronic structure near the gap depends sensitively on the microstructural arrangements of Ag-Sb atoms, suggesting that large {\it ZT} values may originate from the nature of these ordering arrangements. 
\end{abstract}

\maketitle

  Thermoelectrics (TE) drive wide variety of applications related to solid state refrigeration and small scale power generation.  Over last four decades narrow band-gap semiconductor alloys based on Bi-Te compounds for cooling, and Si$_{1-x}$Ge$_x$ and Pb-Te alloys for power generation have dominated technological applications. In recent years there has been a strong impetus to improve the efficiency of TE and this has led to a  constant search of new materials and new concepts.

  Success in discovering novel TE hinges on the ability to achieve the rather challenging task of synthesizing materials with simultaneously high electronic conductivity ({\it $\sigma$}), high thermopower ({\it S}) and low thermal conductivity ({\it $\kappa$}). These solid state properties define the figure of merit {\it ZT=$\sigma$S$^2$T/$\kappa$}, where {\it T} is the operating temperature.  Several new ideas have been proposed to achieve high {\it ZT} values ($>$1). One is the so-called electron crystal phonon glass (ECPG) where one reduces the lattice thermal conductivity without dramatically reducing the power factor ({\it $\sigma$S$^2$}) ~\cite{Slack}.  The other idea is to increase {\it $\sigma$S$^2$} by manipulating electronic density of states (DOS) using 0, 1, and 2 dimensional quantum confinement effects ~\cite{Hicks}. In general {\it $\sigma$S$^2$} depends on the DOS through the transport distribution function $\Sigma_{i}(\epsilon)$ given by:
\begin{equation}
\Sigma_{i}(\epsilon)= \sum_{\vec{k}} v_{i}(\vec{k})^2 \tau (\vec{k}) \delta (\epsilon - \epsilon (\vec{k})) 
\label{eq.1}
\end{equation}  
where {\it i } is the transport direction, the summation is over the first Brillouin zone, $v_{i}(\vec{k})$ is the velocity of the carriers with  wave vector $\vec{k}$, and $\tau (\vec{k})$ is the relaxation time. Yet another concept to increase {\it ZT} by introducing a narrow $\Sigma_{i}(\epsilon)$ centered near the chemical potential was suggested by Mahan and Sofo ~\cite{Mahan}. The last two concepts introduce resonant-like states near the Fermi energy E$_f$.

\begin{figure}[h]
  \centering\includegraphics[scale=0.3]{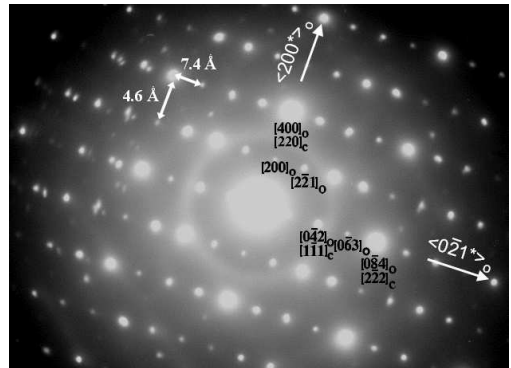}\\[-10pt]
\caption{\label{xraydiff}Electron diffraction pattern showing a lowering of the crystal symmetry from cubic to orthorhombic. It corresponds to [1$\bar{1} \bar{2}$] zone axis in fcc symmetry with cubic unit cell (a=6.4 $\AA$). Superstructure spots can be indexed in orthorhombic symmetry according to a$_0$ $\sim$9.1 $\AA$, b$_0$ $\sim$15.8 $\AA$ and c$_0$ $\sim$22.3 $\AA$ unit cell. The corresponding zone axis is [012]$_0$. Indexation is given for both symmetries. }
\end{figure}

  Recently a new family of complex chalcogenide compounds, \AgSbPbTe\ was described; several members of this family (m=18) when doped appropriately exhibit large {\it ZT} values ($\sim$2.2) at 800K  ~\cite{Kuei}. These compounds were originally designed to have an average cubic  NaCl  structure ({\it Fm$\bar{3}$m} symmetry) with Ag, Pb and Sb atoms being statistically disordered on the Na sites. In fact powder X-ray diffraction studies support this model. However our careful single crystal  X-ray diffraction and electron diffraction studies reveal that is not the case (see Fig.~\ref{xraydiff}). There is clear experimental evidence that the Ag and Sb atoms are not statistically disordered with the Pb atoms. Instead there is a strong driving force that causes long range ordering in the crystal, the nature of which depends on the value of m and the exact experimental conditions with which the materials were prepared ~\cite{Quarez}. In fact  it is possible to create a statistical disorder of all cations on the Pb sites in the samples (such as quenching from a melt) but such samples have inferior TE properties. 

  Given the lower crystallographic symmetry of \AgSbPbTe\ we cannot regard these systems as solid solutions between  AgSbTe$_{2}$ and PbTe but as bona-fide quaternary compounds. This is not surprising since there is a strong enthalpic driving force to not disperse  Ag$^{+}$ and  Sb$^{3+}$ randomly in the structure lest we create coulombic instabilities as we have argued earlier ~\cite{Kuei}. In addition, it is not correct to view the Ag and Sb atoms as dopants because they are present in large stoichiometric proportions. The \AgSbPbTe\ materials achieve a higher { \it ZT} than PbTe at elevated temperatures by exhibiting different temperature dependence in all three properties influencing {\it ZT} ~\cite{Kuei}. Therefore there is a substantial modulating effect on the TE properties of PbTe when the Ag and Sb atoms adopt the proper ordering patterns in it. In this letter we discuss the  electronic structures of several \AgSbPbTe\ compounds (m=10, 16, 18, and 30) based on different ordering models to investigate this issue. The results of our calculations demonstrate that significant qualitative changes occur in the electronic DOS near E$_f$ in \AgSbPbTe\ vis-$\grave{a}$-vis PbTe that could cause favorable enhancements in {\it $\sigma$S$^2$} and its temperature dependence.

  Given the close structural relationship of \AgSbPbTe\ to PbTe, a fundamental question is how the PbTe electronic structure gets modified by extensive substitution with Ag and Sb atoms and their microstructural arrangements. In order to address this we have performed first {\it ab initio} electronic structure calculations using full-potential density-functional method for a series of different microstructural arrangements. We have used the WIEN2K package for our calculations ~\cite{WIEN2K}. This is a linearized full-potential augmented plus local orbitals method within the density functional theory formalism ~\cite{DFT}. We took the generalized gradient approximation in Ref.~\cite{GGA} for the exchange and correlation potential. Scalar relativistic corrections were included and spin-orbit interaction was incorporated  using a second variational procedure ~\cite{Koelling}. Convergence of the self-consistent iterations was performed using 10 (for single and Ag-Sb pair atoms$^{\dag}$), 15 (for layer of Ag-Sb pairs), 18 (for chain of Ag-Sb pairs) and 2 (for cluster of Ag-Sb pairs) $\vec{k}$ points inside the reduced Brillouin zones to within 0.0001 Ry with a cutoff of -6.0 Ry between the valence and the core states.

\begin{figure}[h]
  \centering\includegraphics[scale=0.23]{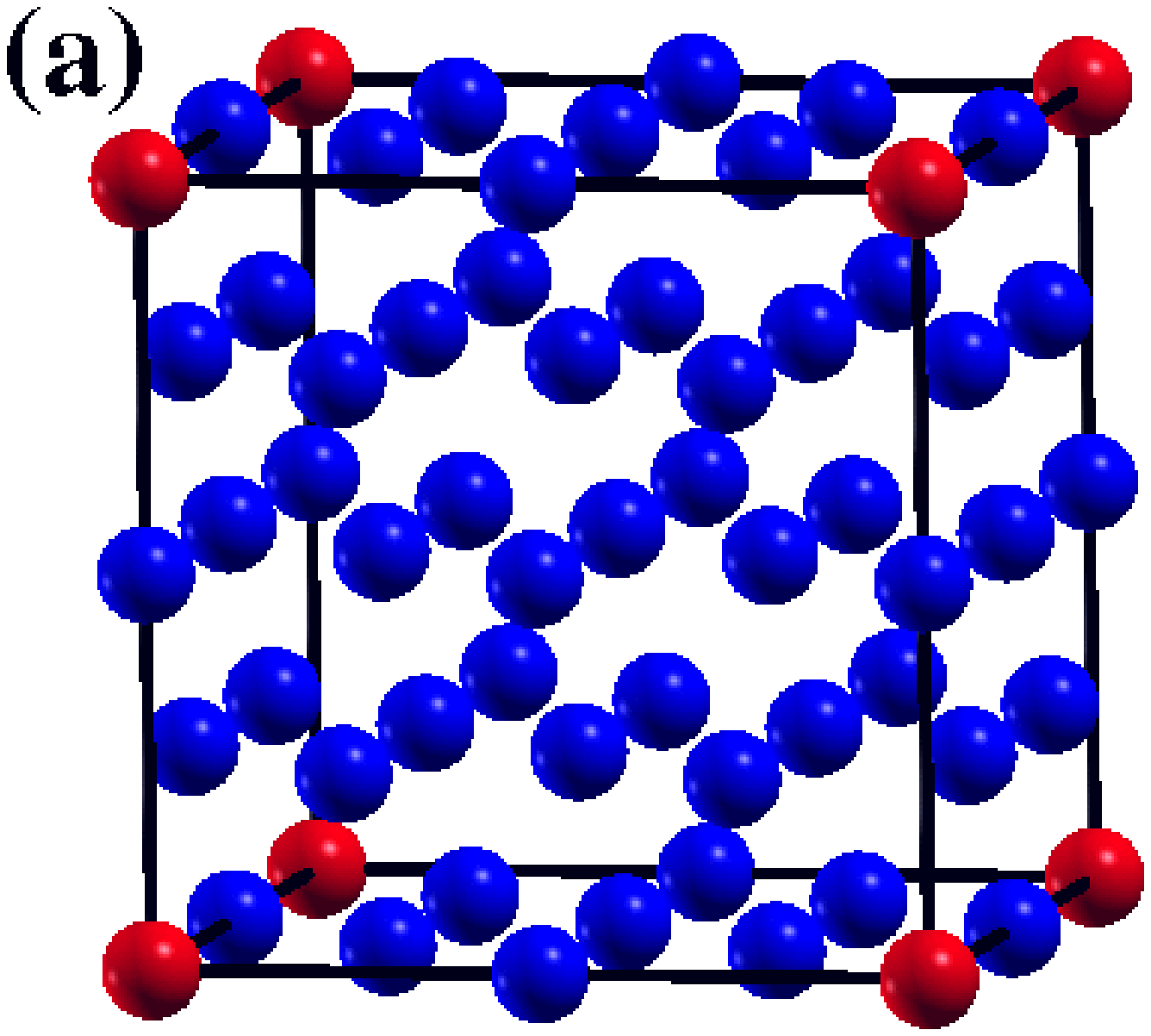}
  \centering\includegraphics[scale=0.23]{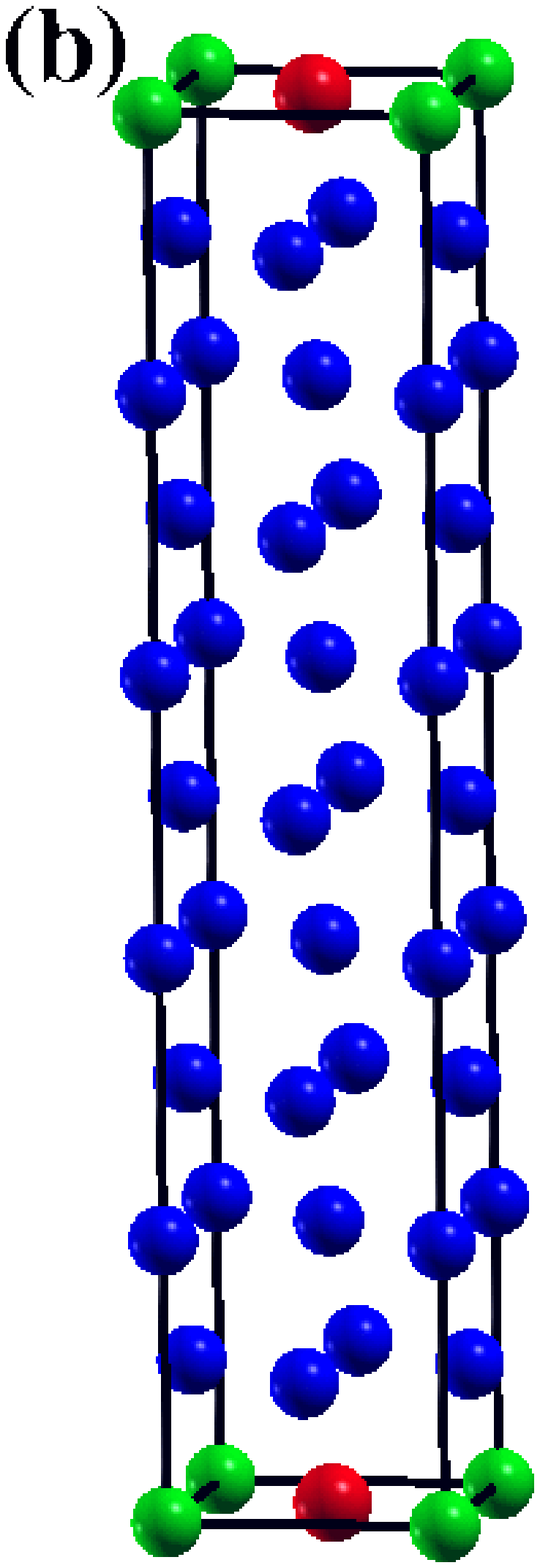}
  \centering\includegraphics[scale=0.23]{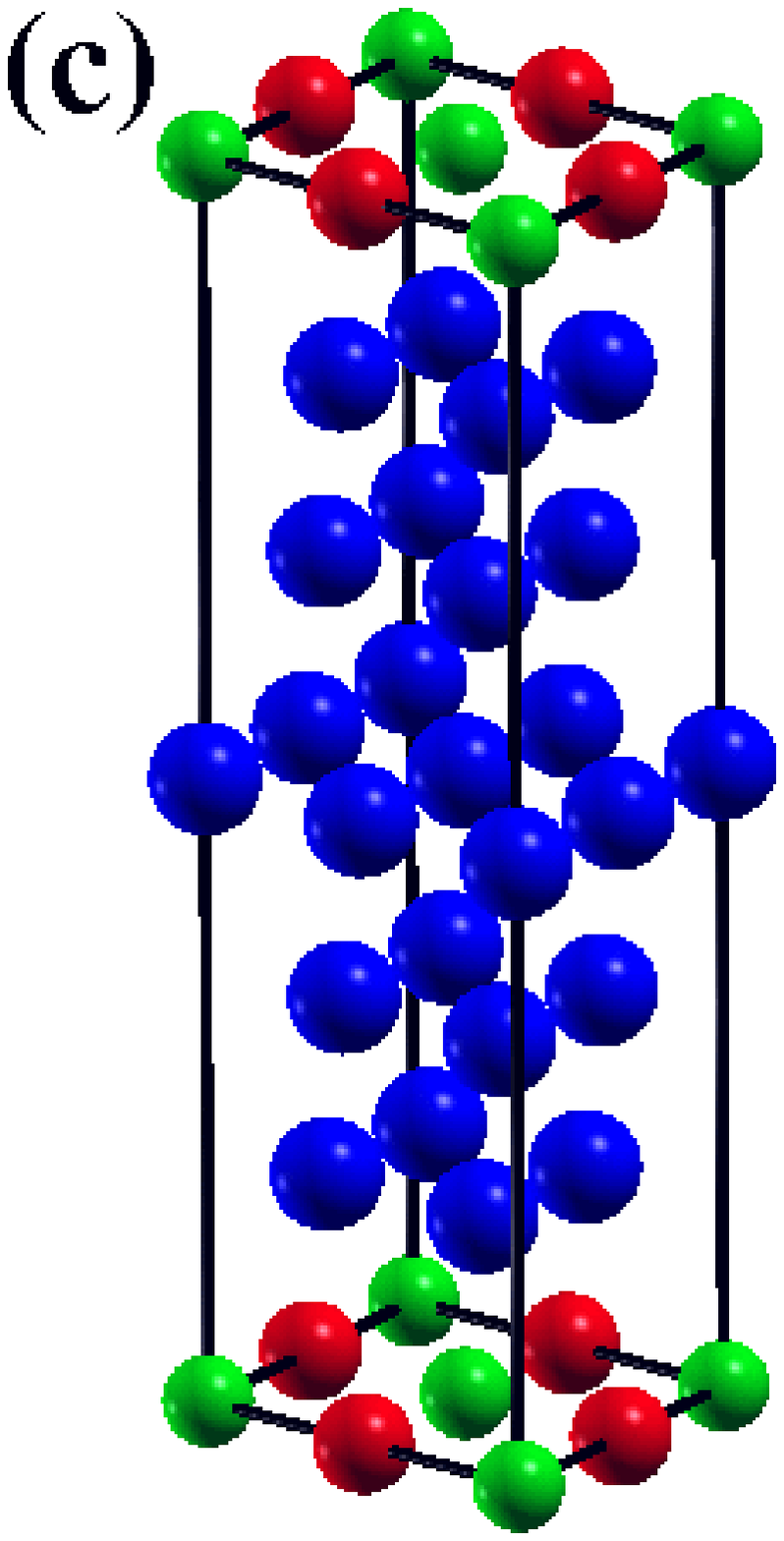}
  \centering\includegraphics[scale=0.23]{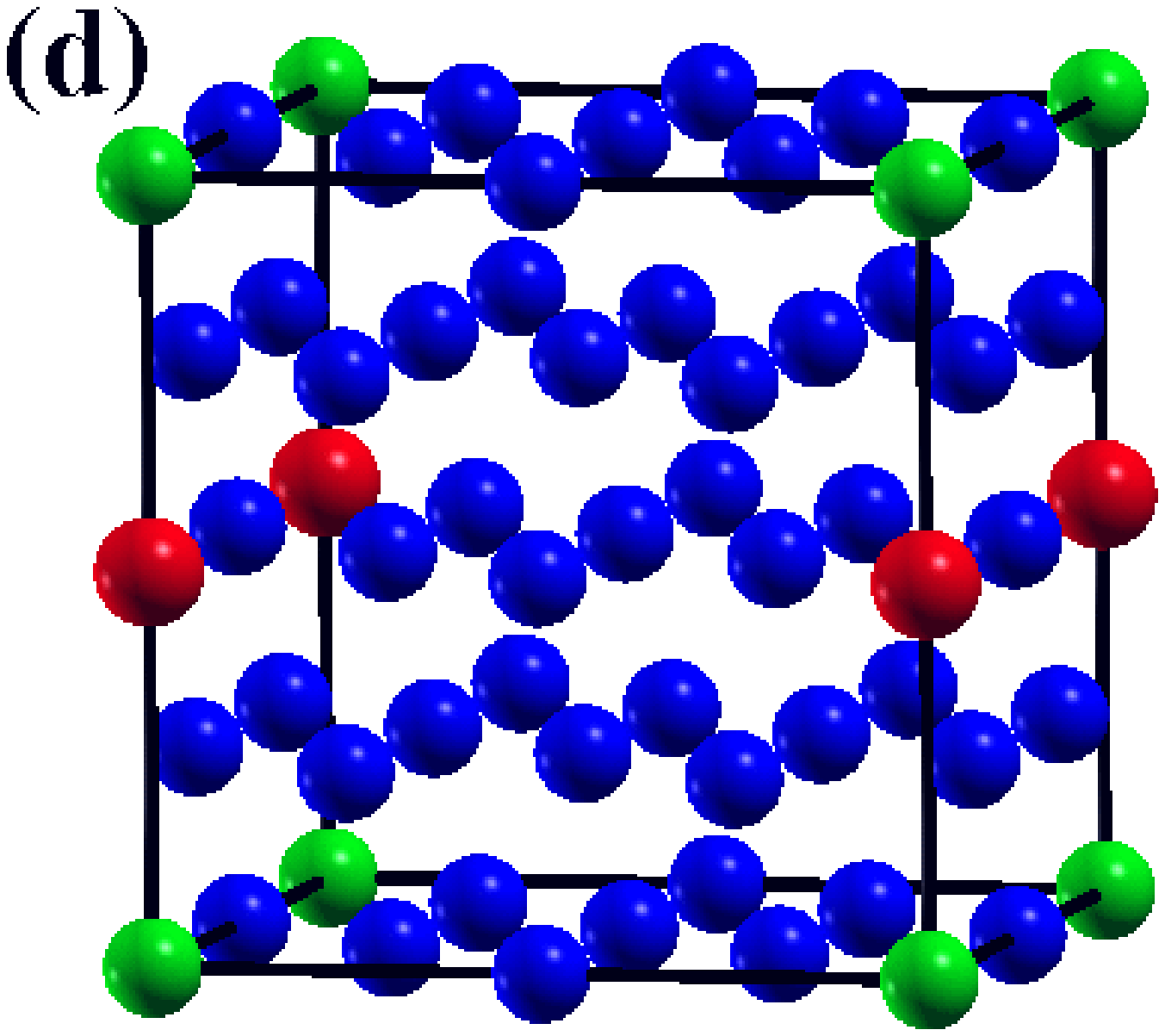}
  \centering\includegraphics[scale=0.23]{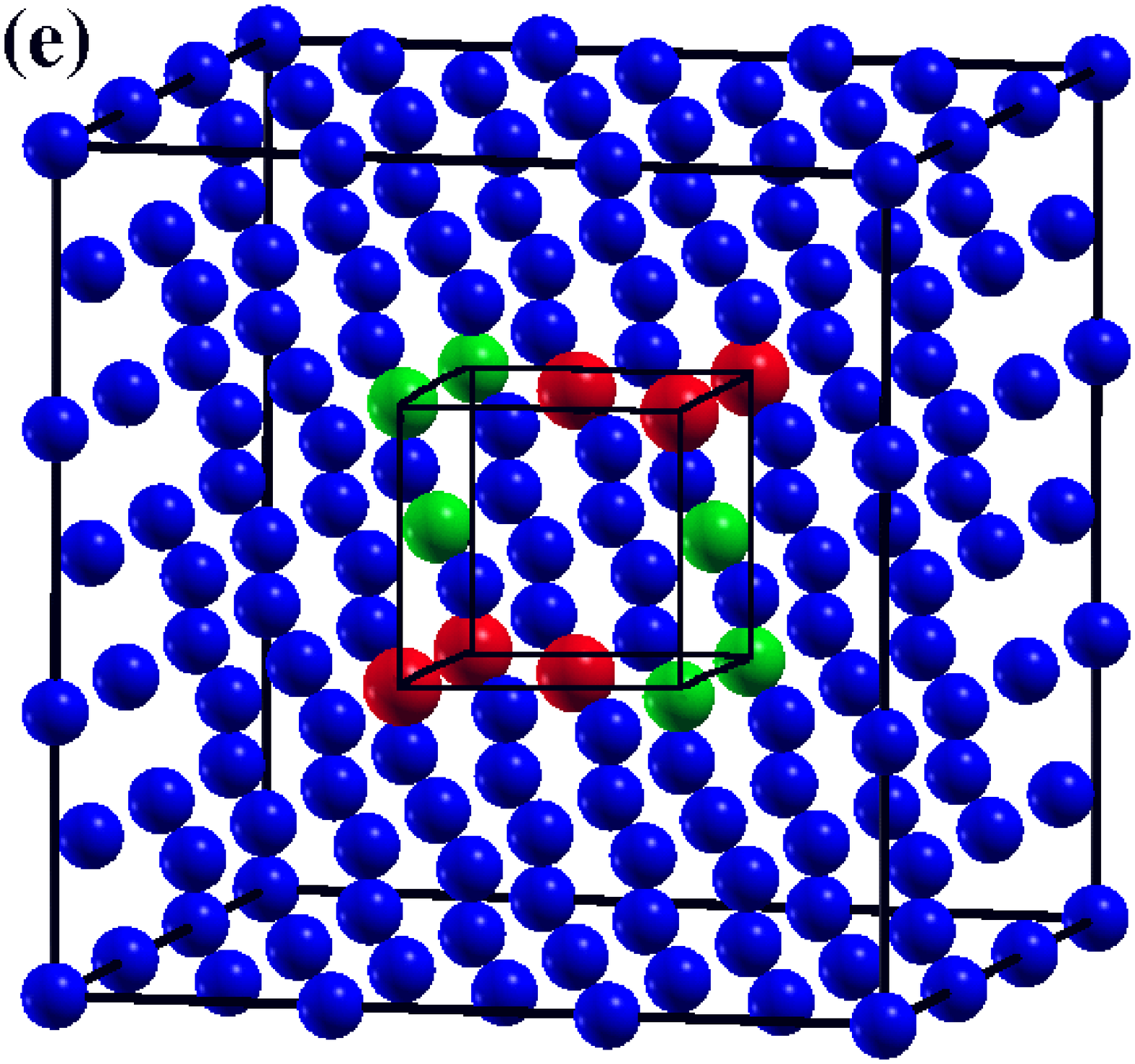}\\[-10pt]
\caption{\label{struct}Unit cell models for (a) single Ag atom in AgPb$_{31}$Te$_{32}$ , (b) Ag-Sb layer perpendicular to [001] direction in AgSbPb$_{18}$Te$_{20}$, (c)  Ag-Sb layer perpendicular to the fcc [111] direction in AgSbPb$_{10}$Te$_{12}$, (d) Ag-Sb chain parallel to the [001] direction in AgSbPb$_{30}$Te$_{32}$, and (e) Ag-Sb cluster in AgSbPb$_{16}$Te$_{18}$. For the reason of clarity we show only Pb fcc lattices with Pb in blue, Ag in red, and Sb in green colors.}
\end{figure}

  First, we performed calculations with isolated Ag and Sb atoms in the PbTe lattice in order to obtain a clear picture of their individual role in modifying the electronic structure of PbTe (Fig.~\ref{struct}a). Then both Ag and Sb were introduced to simulate stoichiometries relevant to those of \AgSbPbTe\ compounds. Given that the exact crystal structure is not known, several plausible microstructural models were examined, all of which involved long range ordering of the atoms. In one model the Ag and Sb atoms were placed in monolayers (Fig.~\ref{struct}b,c). In another model the Ag and Sb atoms were placed along straight infinite chains running parallel to a crystallographic unit cell axis (e.g. c-axis)(Fig.~\ref{struct}d). In yet a third arrangement, the atoms were placed in the center of a 3x3x3 supercell to create a ``AgSbTe$_2$ nanodot'' embedded in a PbTe matrix (Fig.~\ref{struct}e). The chain and cluster models are in qualitative agreement with our experimental transmission electron microscopy (TEM) observations of Ag-Sb ordering ~\cite{Kuei}. Although we believe these arrangements capture much of the crystal physics in these materials, we recognize that there are many more that could be considered which nevertheless should result in similar general conclusions.

  To model the Ag(Sb) isolated and Ag-Sb pair atoms  we constructed  2x2x2 supercells with 64 atoms. For the isolated case, we chose Ag(Sb) at the origin of the supercell with a separation of two lattice constants (12.924 $\AA$) between the Ag(Sb) atoms (Fig.~\ref{struct}a). For the Ag-Sb pair, we considered two arrangements (not shown in Fig.~\ref{struct}), one where the Ag and Sb are far apart (Sb at the origin and Ag at the center of the supercell) with a separation distance of $\sim$11.19 $\AA$ and the other where the Ag and Sb are as close as possible (Sb at the origin and Ag at the next nearest neighbour site of Sb) with a separation distance of $\sim$4.57 $\AA$. For the structure where Ag-Sb layers are separated by several Pb layers we also considered two cases where the Ag-Sb layer is normal to the [001] direction in a 1x1x5 supercell (40 atoms/cell) with the Ag-Sb layer located in the z=0 plane (Fig.~\ref{struct}b) and where the Ag-Sb layer is normal to the [111] direction (Fig.~\ref{struct}c). The fcc unit cell can be viewed along the [111] direction as a hexagonal unit cell. We have used a 2x2x2 hexagonal supercell (48 atoms/cell) with the Ag-Sb layer perpendicular to the c-axis which is the [111] direction in the fcc unit cell. To model the chains we used a 2x2x2 supercell where the Ag-Sb chains are oriented parallel to the [001] direction (Fig.~\ref{struct}d) and separated by 12.924 $\AA$. Finally, for the ``AgSbTe$_2$'' clusters we constructed a 3x3x3 supercell (216 atoms/cell) and the cluster consists of six Ag-Sb pairs located at the center of the supercell with a minimum separation distance between two clusters of $\sim$12.924 $\AA$ (Fig.~\ref{struct}e).

  The nature of defect states in narrow band gap semiconductors in general and PbTe in particular are important problems but still not completely understood.  Lent { \it et. al.}~\cite{Lent} argued that in PbTe, due to its large dielectric constant Coulomb effects associated with charged impurities were screened out and local bonding effects determined the nature of the defect states. Instead of donor or acceptor states appearing in the band gap as seen in conventional wide band gap semiconductors they argued that one should see resonance states outside of the band gap region. In this letter we provide for the first time theoretical justification of this idea starting from first principles. As discussed below, our {\it ab initio} calculations clearly show that the states associated with isolated Ag(Sb) atoms in PbTe are not only resonant with the valence band (VB) and conduction band (CB) states of PbTe but they also reduce the band gap.
 
\begin{figure}[b]
 \centering\includegraphics[scale=0.20, angle=-90]{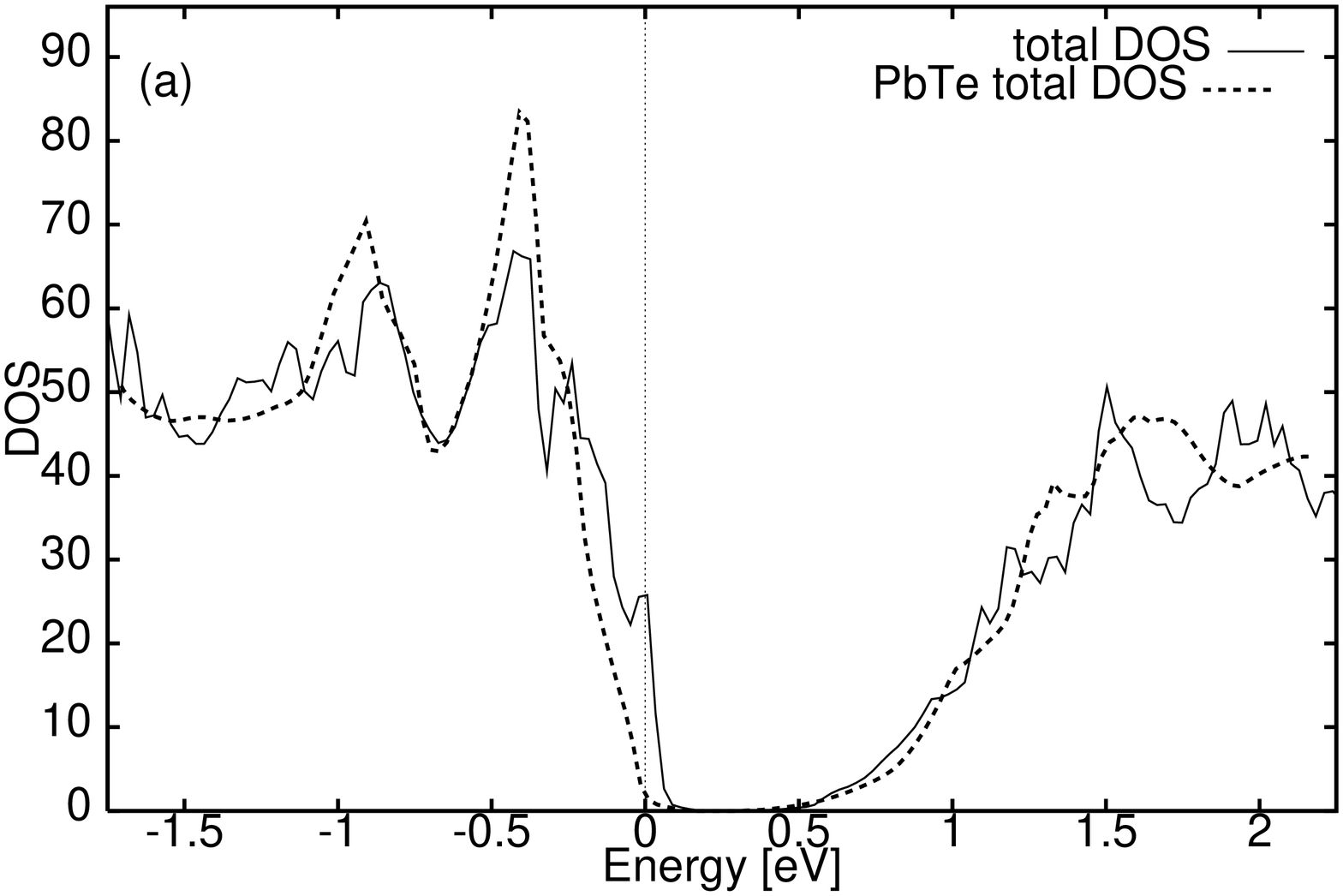}
  \centering\includegraphics[scale=0.20, angle=-90]{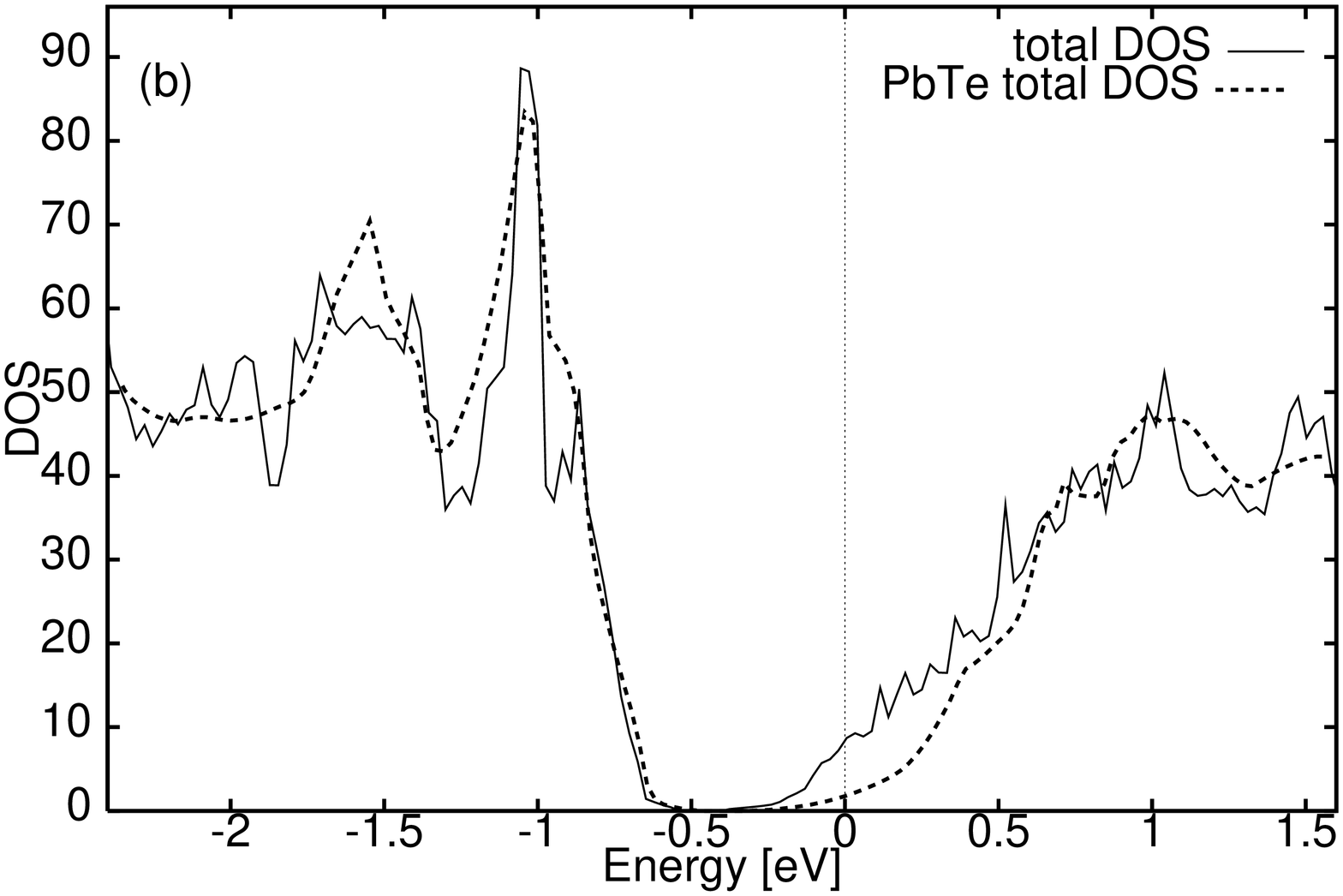}
  \centering\includegraphics[scale=0.20, angle=-90]{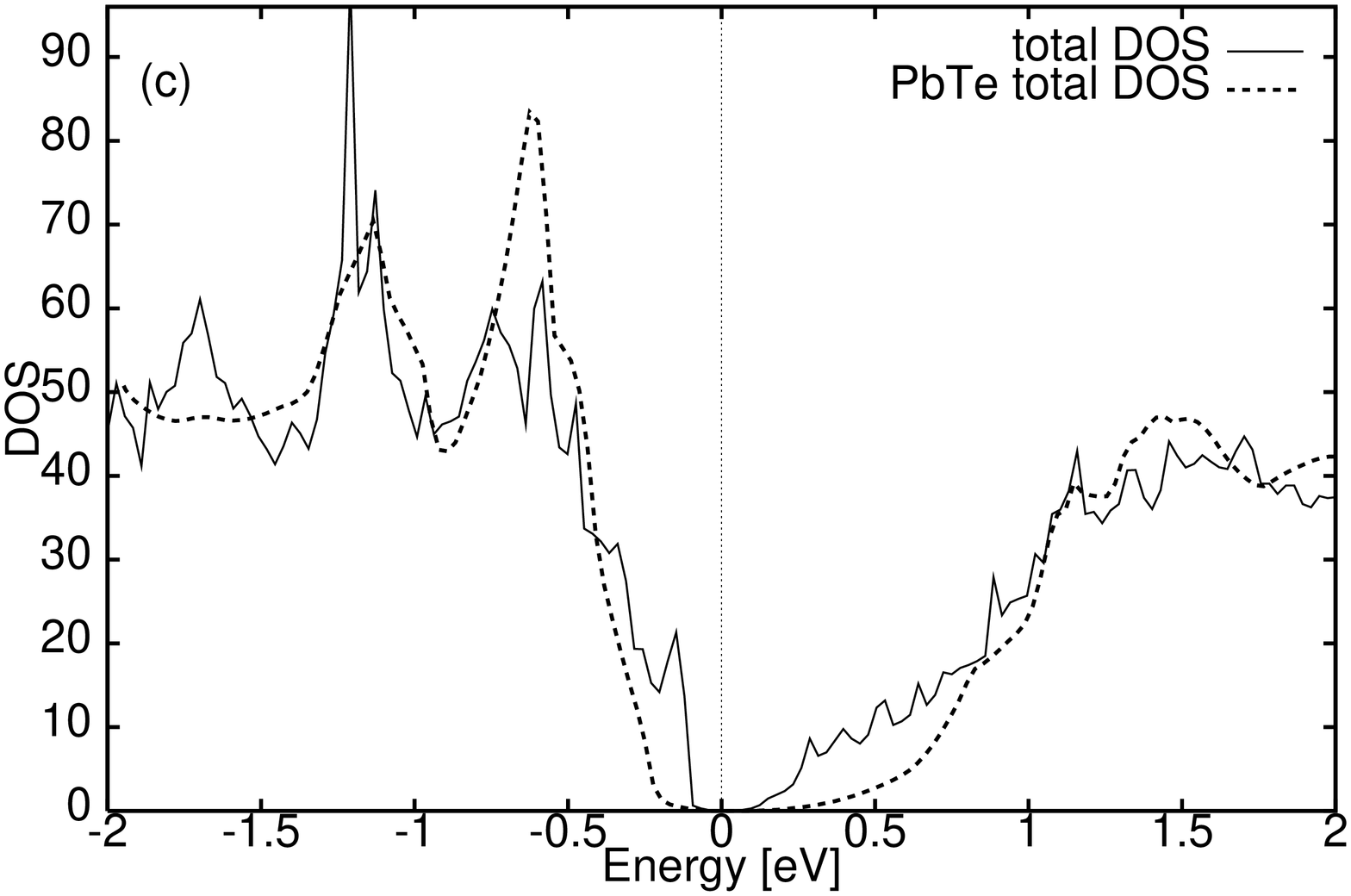}
  \centering\includegraphics[scale=0.20, angle=-90]{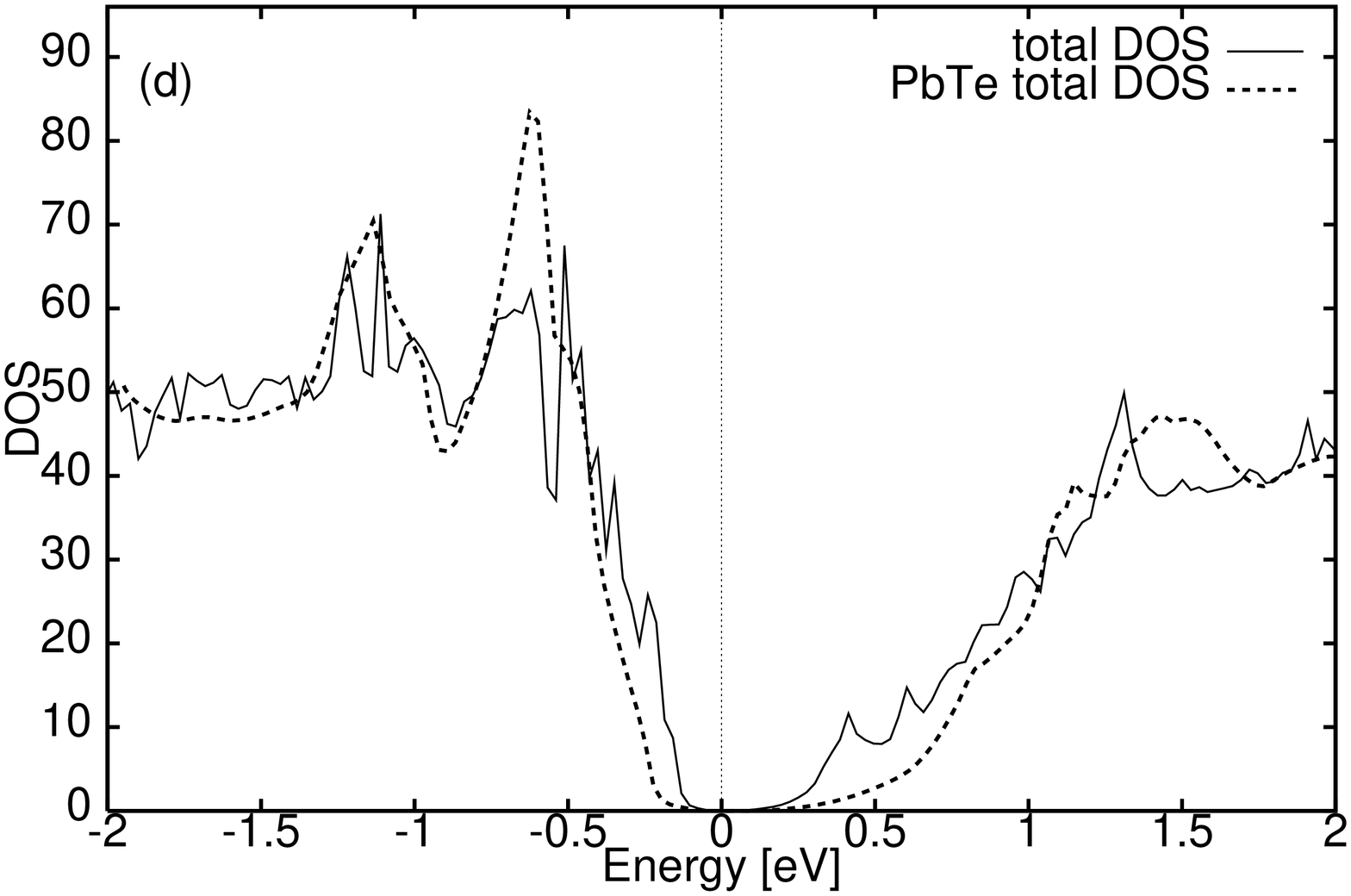}\\[-10pt]
 \caption{\label{singleimpDOS}Total DOS of (a) single Ag atom, (b) single Sb atom, (c)  Ag-Sb pair with the Ag-Sb distance of $\sim$11.19 $\AA$, and (d) Ag-Sb pair with Ag-Sb distance of $\sim$4.57 $\AA$, in PbTe. For comparison the total DOS of PbTe is shown in dashed line.}
 \end{figure}

  The total DOS for isolated Ag atoms is shown in Fig.~\ref{singleimpDOS}a. It can be seen that Ag introduces states near the top of the PbTe VB. Partial DOS analysis shows that these states consist mostly of p orbitals of the six nearest neighbor Te atoms of Ag. These states are resonant with the VB and extend into the PbTe gap region.  On the other hand the isolated Sb single atoms introduce resonant states near the bottom of the PbTe CB (Fig.~\ref{singleimpDOS}b) which  extend nearly $\sim$0.75 eV into the CB starting from the CB bottom. Sb and its Te nearest neighbours atoms have the highest contribution to these resonant states. The Sb p states hybridize with Te p states in the range (-0.25,0.5) eV. Therefore these states are not only resonant with the PbTe CB but they extend into the PbTe gap. 

  Results for isolated  Ag-Sb pairs are consistent with the Ag(Sb) single atom results in the sense that Ag introduces new states near the top of VB, whereas Sb introduces new states near the bottom of CB (Fig.~\ref{singleimpDOS}c,d) decreasing the PbTe gap. Both cases (Ag-Sb far apart and Ag-Sb next nearest neighbours)(not shown in Fig.~\ref{struct}) show semiconducting behaviour with a very small gap and a more rapidly increasing DOS near the VB and CB extrema as compared to the DOS of PbTe. The specific features of the DOS in the gap region are very different for these two cases. Total energy comparison for AgSbPb$_{30}$Te$_{32}$ shows that these two structures are very close in energy, the case when Ag-Sb atoms are far apart has a lower energy by $\sim$20 meV/(unit cell).

\begin{widetext}
\begin{figure}[t]
   \begin{minipage}[b]{1.0\textwidth}
   \begin{minipage}[b]{.5\textwidth}
  \centering\includegraphics[scale=0.21, angle=-90]{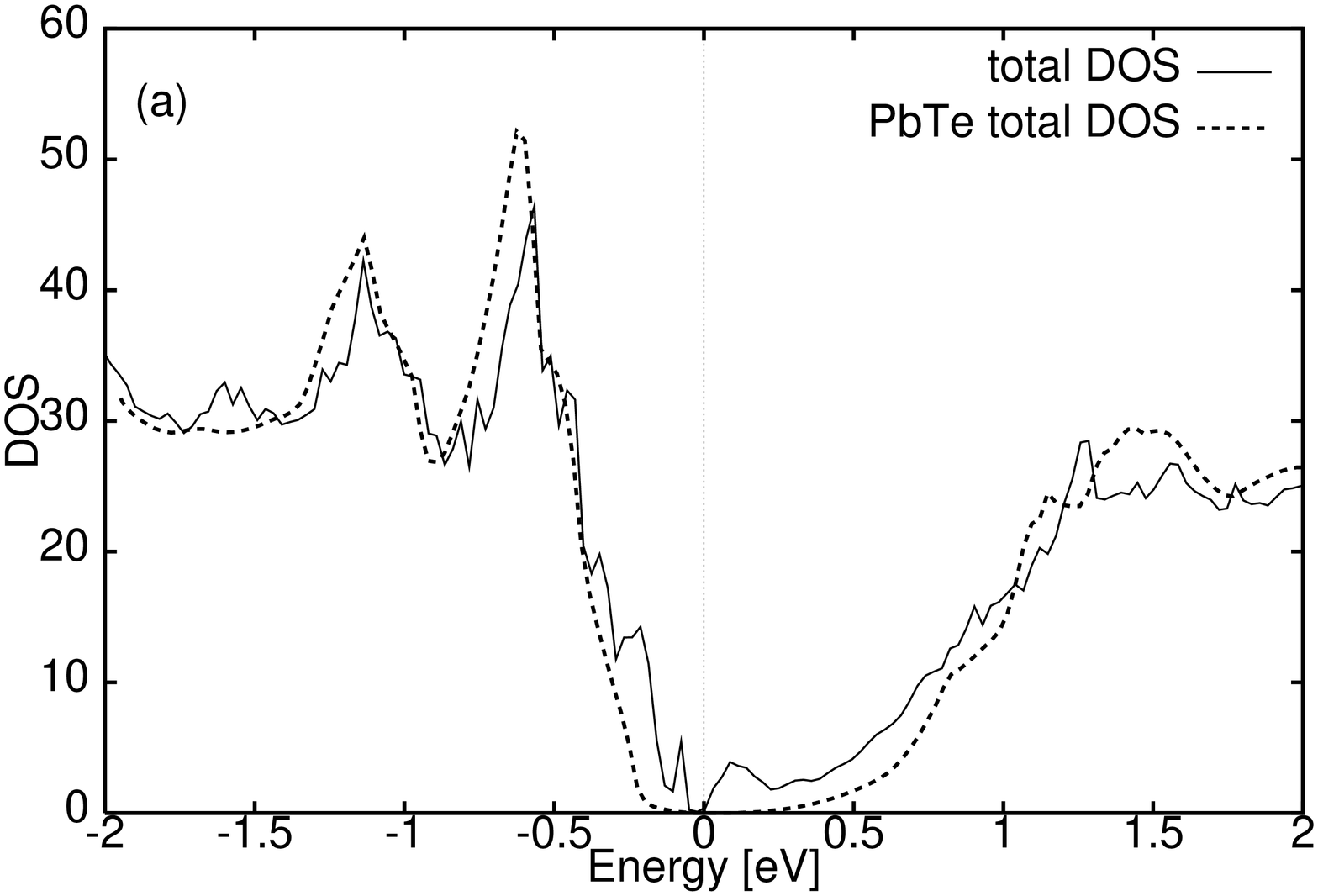}
  \centering\includegraphics[scale=0.21, angle=-90]{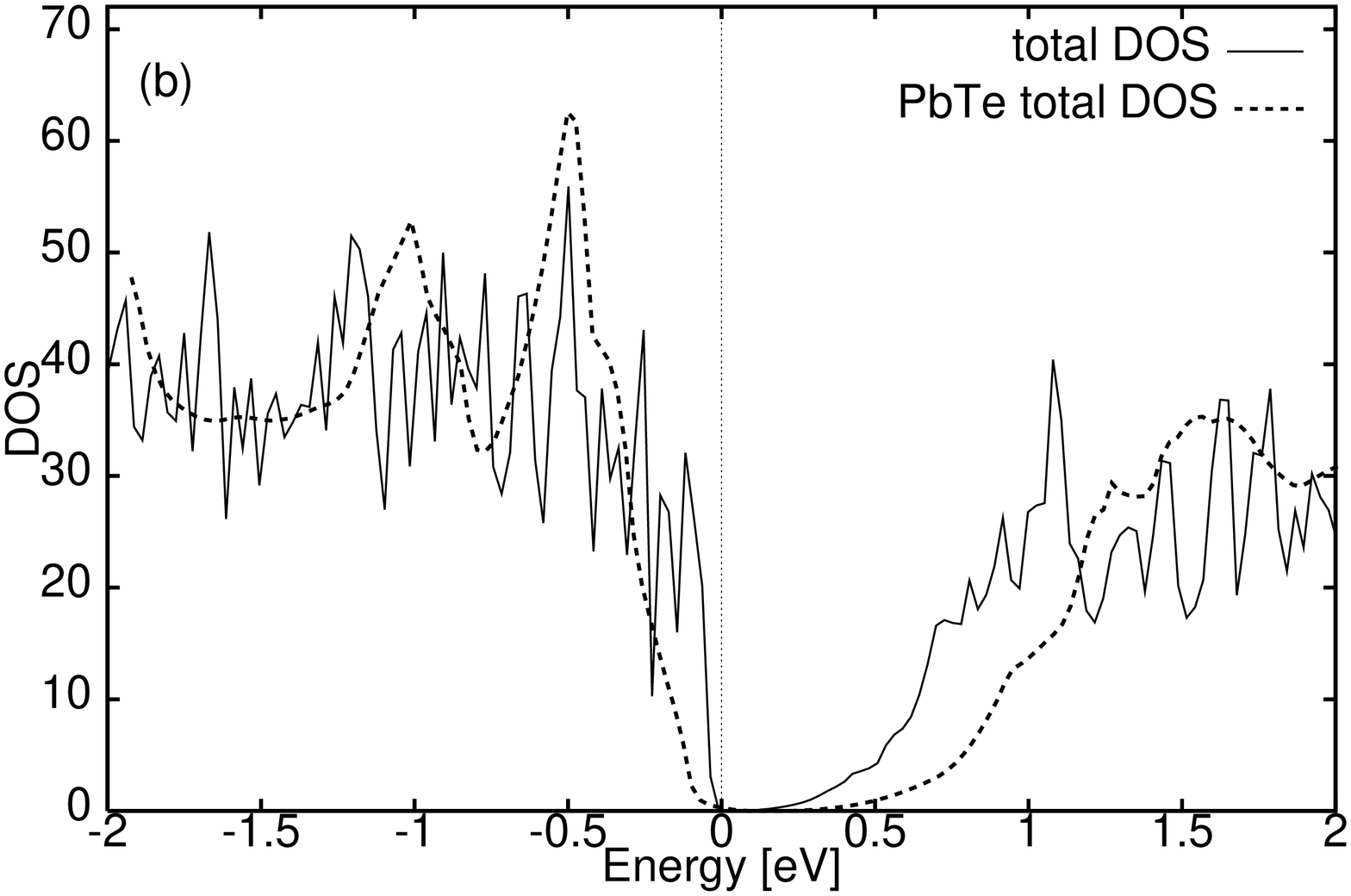}
  \end{minipage}%
   \begin{minipage}[b]{.5\textwidth}
  \centering\includegraphics[scale=0.21, angle=-90]{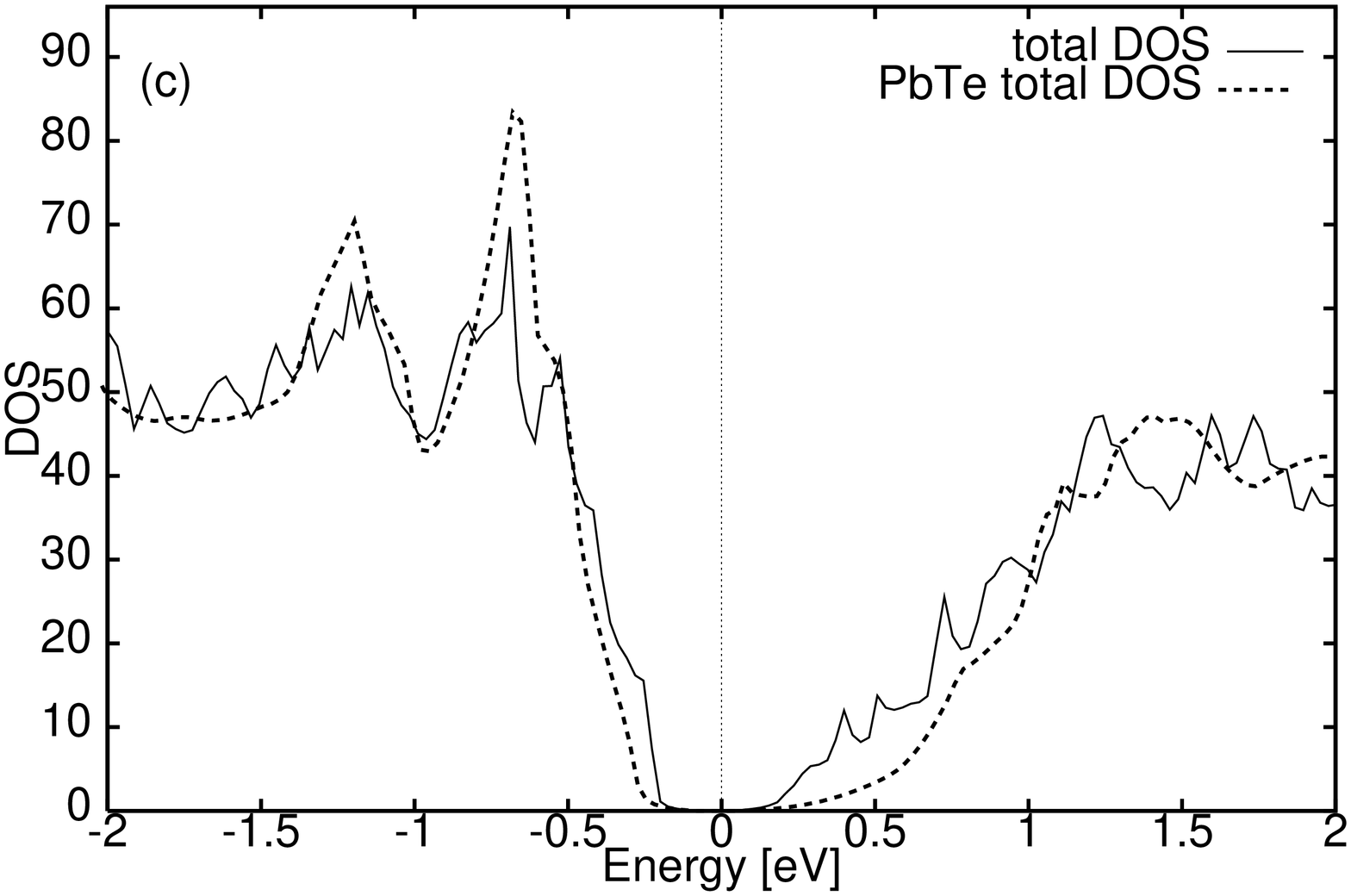}
  \centering\includegraphics[scale=0.21, angle=-90]{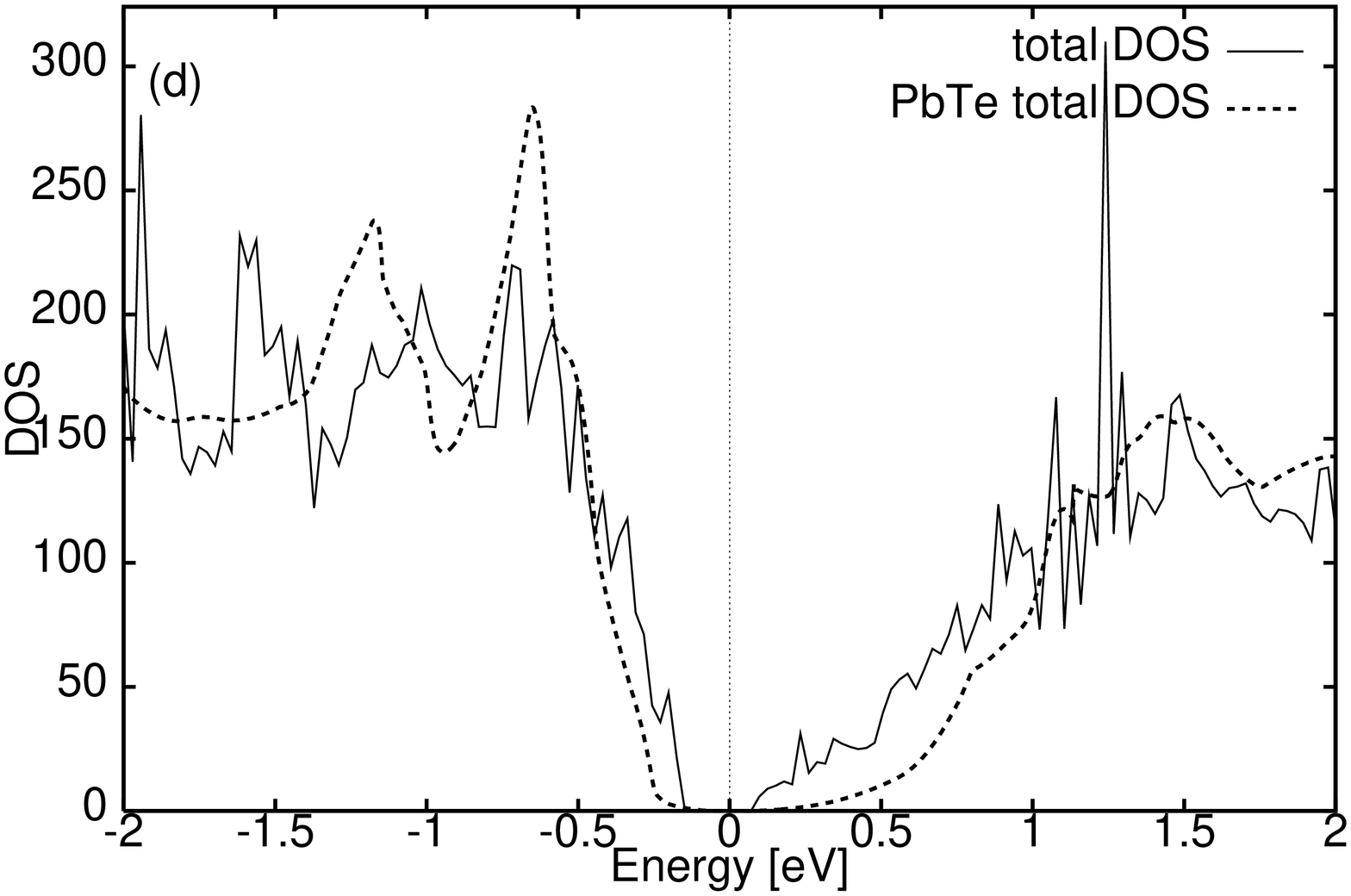}
  \end{minipage}
  \caption{\label{microstructDOS}Total DOS of (a) Ag-Sb layer model perpendicular to [001] direction, (b) Ag-Sb layer model perpendicular to [111] direction, (c)  Ag-Sb chain model, and (d) Ag-Sb cluster model. The total DOS of PbTe is shown in dashed line.}
  \end{minipage}
 \end{figure}
\end{widetext}

  It is interesting to compare the DOS results for the layer structures of Ag-Sb (Fig.~\ref{struct}b,c). When the Ag-Sb layer is perpendicular to the [001] direction, the states associated with the Ag-Sb layer completely fill the PbTe gap giving a semimetallic behaviour (Fig.~\ref{microstructDOS}a), whereas the Ag-Sb  layer perpendicular to [111] direction show a semiconducting behaviour (Fig.~\ref{microstructDOS}b). This indicates that the electronic structure of \AgSbPbTe\ systems and consequently the electronic properties are very sensitive to the microstructural arrangements of Ag-Sb atoms. The Ag-Sb chain model shows semiconducting behaviour (Fig.~\ref{microstructDOS}c). This chain model has the same stoichiometry (AgSbPb$_{30}$Te$_{32}$) as the Ag-Sb pair models. Total energy comparisons show that the chain model has a lower energy by 0.2 eV/(unit cell) than the Ag-Sb pair models suggesting that Ag-Sb chain orderings along [001] directions are favorable microstructures.  This is consistent with the results of electron crystallographic studies which indicate the presence of Ag-Sb chains in the crystal ~\cite{Quarez}. The DOS results for the ``AgSbTe$_2$'' cluster model show also semiconductor behaviour (Fig.~\ref{microstructDOS}d).
 \\[-10pt]

  The calculations for different microstructural arrangements of Ag-Sb in PbTe show a generic feature; when Sb atoms replace Pb atoms, Sb hybridize with Te atoms forming strong covalent interactions. When Ag replace Pb, the p states of Te which are the nearest neighbours of Ag are strongly perturbed. Therefore the electronic structure of \AgSbPbTe\ compounds depend sensitively of these perturbed Te states by the Ag(Sb) atoms.    

  We have shown for the first time that the details of the DOS near the energy gap of \AgSbPbTe\ depend sensitively on the microstructural ordering arrangements of Ag-Sb pairs in PbTe. The nature of the states near the top VB and bottom CB in these quaternary compounds is substantially different than in PbTe. The common feature of these Ag-Sb arrangements is that they have a more rapidly increasing DOS near the gap as compared to bulk PbTe due to the appearance of distinct resonant states. It is well accepted that resonant structures in the DOS near E$_f$, created by quantum size effects ~\cite{Hicks}~\cite{Harman}, superlattice engineering ~\cite{Rama} or chemical means ~\cite{Mahan}~\cite{Kanatzidis} are very desirable features because they could enhance the TE figure of merit {\it ZT}.
  
  Financial support from the Office of Naval Research (Contract No. N00014-02-1-0867 MURI program) is greatfully acknowledged.

$^{\dag}$  The terms ``Ag-Sb pairs'' and ``Ag-Sb chains'' are used for convenience and should not be confused with actual pairs or chains possessing Ag-Sb bonds. It is implicit that only the Te atoms are bonded to the metals and serve as bridges e.g. Ag-Te-Sb.


\begin{thebibliography}{99}

\bibitem{Slack}G. A. Slack, {\it Solid State Physics}, Vol.{\bf 34}, 1 (1979); G. S. Nolas, J. Sharp, and H. J. Goldsmid, {\it Thermoelectrics: Basic Principles and New Materials Developments}, p 177, Springer, New York, (2001).
\bibitem{Hicks}L. D. Hicks, and M. S. Dresselhaus, \Journal{ Phys. Rev.}{B 47}{12727 }{1993}.
\bibitem{Mahan}G. D. Mahan, and J. O. Sofo,\Journal{Proc. Natl. Acad. Sci.}{93}{7436}{1996}. 
\bibitem{Kuei}K. F. Hsu, S. Loo, F. Guo, W. Chen, J. S. Dyck, C. Uher, T. Hogan, E. K. Polychroniadis, and M. G. Kanatzidis, \Journal{ Science}{303}{818}{2004}.\bibitem{Quarez}E. Quarez, K. F. Hsu, R. Pcionek, and M. G. Kanatzidis, manuscript in preparation.   
\bibitem{WIEN2K}P. Blaha {\it et al.}, WIEN2K, An Augmented Plane Wave + Local Orbitals Program for Calculating Crystal Properties, Karlheinz Schwarz, Techn. Universitat Wien, Austria, 2001.
\bibitem{DFT}P. Hohenberg, and W. Kohn, \Journal{ Phys. Rev.}{136}{B864}{1964}; W. Kohn, and L. Sham, \Journal{ibid}{140}{A1133}{1965}.
\bibitem{GGA}J. P. Perdew, K. Burke, and M. Ernzerhof, \Journal{ Phys. Rev. Lett.}{77}{3865}{1996}.
\bibitem{Koelling}D. D. Koelling, and B. Harmon, \Journal{J. Phys.}{C 13}{6147}{1980}.
\bibitem{Lent} C. S. Lent, M. A. Bowen, R. S. Allgaier, J. D. Dow, O. F. Sankey, and E. S. Ho, \Journal{Solid State Commun}{61}{83}{1987}. 
\bibitem{Harman} T. C. Harman, P. J. Taylor, M. P. Walsh, and B. E. LaForge, \Journal{Science}{297}{2229}{2002}; L. D. Hicks, T. C. Harman, and M. S. Dresselhaus, \Journal{Appl. Phys. Lett.}{63}{3230}{1993}.
\bibitem{Rama} R. Venkatasubramanian, E. Siivola, T. Colpitts, and B. O'Quinn, \Journal{Nature}{413}{597}{2001}.
\bibitem{Kanatzidis}M. G. Kanatzidis, { \it Semiconductors and Semimetals}, { \bf 69}, 51 (2001); { \it Chemistry, Physics, and Material Science of Termoelectric Materials: Beyond Bismuth Telluride}, p 35, Kluwer Academic, New York, (2003). 


\end{thebibliography}
\end{document}